*Agent-Based Recovery Model for Seismic Resilience Evaluation of Electrified Communities*

**Li Sun[1]∗; Bozidar Stojadinovic[1]; Giovanni Sansavini[2]**

[1]Chair of Structural Dynamics and Earthquake Engineering, Institute of Structural Engineering, ETH Zurich, Switzerland

[2]Laboratory of Reliability and Risk Engineering, Institute of Energy Technology, ETH Zurich, Switzerland

**ABSTRACT:** In this paper, an agent-based framework to quantify the seismic resilience of an Electric Power Supply System (EPSS) and the community it serves is presented. Within the framework, the loss and restoration of the EPSS power generation and delivery capacity and of the power demand from the served community are used to assess the electric power deficit during the damage absorption and recovery processes. Damage to the components of the EPSS and of the community built environment is evaluated using the seismic fragility functions. The restoration of the community electric power demand is evaluated using the seismic recovery functions. However, the post-earthquake EPSS recovery process is modeled using an agent-based model with two agents, the EPSS Operator and the Community Administrator. The resilience of the EPSS-Community system is quantified using direct, EPSS-related, and societal, community-related, indicators. Parametric studies are carried out to quantify the influence of different seismic hazard scenarios, agent characteristics, and power dispatch strategies on the EPSS-Community seismic resilience. The use of the agent-based modeling framework enabled a rational formulation of the post-earthquake recovery phase and highlighted the interaction between the EPSS and the community in the recovery process not quantified in resilience models developed to date. Furthermore, it shows that the resilience of different community sectors can be enhanced by different power dispatch strategies. The proposed agent-based EPSS-Community system resilience quantification framework can be used to develop better community and infrastructure system risk governance policies.

∗Corresponding author
Email address: sun@ibk.baug.ethz.ch (Li SUN),
Preprint, on *Risk Analysis*

# 1. INTRODUCTION

The Electric Power Supply System (EPSS) is the backbone of modern communities. Its resilience is crucial for the recovery of the communities affected by natural disasters (Mieler, Stojadinovic, Budnitz, Comerio, and Mahin, 2014). Resilience refers to the capability of a system to decrease the initial damage it may suffer due to disruptive events, together with the ability to bounce back to the appropriate and stabilized functionality level thereafter (Society of Risk Analysis, 2015). Nevertheless, as exemplified by many occurrences in recent history, EPSSs did not prove sufficiently robust to and recoverable from the impact of natural hazards and random technological failures, resulting in serious economic and societal losses (Andersson et al., 2005, Liu et al., 2012, Fujisaki et al., 2014, Kwasinski et al., 2014). In addition, seemingly inconsequential local damage can propagate through EPSSs due to inherent interconnectivity and lead to wide-scale cascading failures, which can further affect other interdependent critical infrastructures (Dueñas-Osorio and Vemuru, 2009, Buldyrev et al., 2010, Zio and Sansavini, 2011). In many cases, the losses after a natural disaster tend to be more severe than expected due to inadequate preparedness and dependence on other infrastructure systems that also impacted by the disaster (Adachi and Ellingwood, 2008, Hollnagel and Fujita, 2013, Yu et al., 2015).

Against this backdrop, the focus of both researchers and practitioners expanded from disaster risk to post-disaster resilience of civil infrastructure systems. It is expected that resilient systems will be robust to catastrophic disruptive events while also able to recover quickly from the initial damage they sustained (Lundberg and Johansson, 2015), thus reducing the risks such systems pose.

The notion of disaster resilience as a time-varying process (Michel-Kerjan, 2015) has recently been extensively examined. There are many proposals to quantify the seismic resilience of civil infrastructure systems and communities (Hosseini et al., 2016). Bruneau et al. (2003) proposed a conceptual framework to consider the seismic resilience of infrastructure systems from four aspects, which are technical, organizational, social and economic. Ouyang et al. (2012) put forward a quantitative model for assessing the total functionality loss of the affected infrastructure systems during the shock absorption and the recovery phases.

Resilience quantification frameworks developed to date address the ability of the civil infrastructure systems to operate and provide service, i.e. to supply the community. The demand of the community for the functions provided by the supporting civil infrastructure systems is assumed to remain unchanged, which is usually not true after natural disasters, particularly earthquakes. The gross functionality loss and, therefore, the disaster risk are difficult to evaluate as the system functionality level after the disaster recovery process will not necessarily be the same as the pre-disaster level (Linkov et al., 2014).

To address these challenges, a compositional resilience quantification framework was developed by Déle and Didier (2014), and applied to quantify the seismic resilience of EPSSs by Sun et al. (2015a) and Didier et al. (2015). In this framework, both the power supply capacity of EPSS and the power demand from the community that it serves are considered simultaneously and tracked as they evolve through the earthquake damage absorption and post-earthquake recovery phases. The gap between the supply and the demand is used to quantify the lack of resilience of the EPSS-Community system. Two measures for such lack of resilience, one that quantifies the electric power deficit/functionality loss, and is directly related to the function of the EPSS, and the other that quantifies the number of people without electric power and is related to the function of the community, are proposed.

The initial performance level losses of each individual component of EPSS (e.g. transformers, switches, and circuit breakers) and the built environment it supplies (e.g. office and apartment buildings, factories, schools, hospitals) in the earthquake damage absorption phase are assessed using seismic Vulnerability Functions (VFs). Seismic VFs describe the probability that the loss of functionality of a component will exceed a given threshold, conditioned on a measure of intensity of the ground motion excitation that the component experienced. Loss of functionality assessment starts by determining the damage state of the components of the community built environment and the civil infrastructure systems. Libraries of seismic damage fragility functions are available for the components of the community built environment (Rossetto and Elnashai, 2003, Kwon and Elnashai, 2006, Jeong and Elnashai, 2007, Senel and Kayhan, 2009, ATC-58, 2015, OpenQuake, 2015, Syner-G, 2015) and civil infrastructure systems (HAZUS, 2015, Syner-G, 2015), including the EPSS components. Once the damage states of each component are determined, their level of functionality is quantified using deterministic relations between damage states and portions of remaining functionality (be they on the supply or on the demand side) proposed by Déle and Didier (2014) to obtain the VFs for the components. System-level vulnerability is computed by aggregating the remaining functionality of the components immediately after the earthquake event. This concerns both the demand and the supply of the electric power and involves a model of EPSS system operation, the so-called EPSS dispatch.

The community and the EPSS will enter the post-earthquake recovery phase after the damage has been absorbed. Recovery Functions (RFs) are developed to represent the recovery process (HAZUS, 2015) of the components of the EPSS-Community system. RFs quantify the probability that the component functionality exceeds a threshold after an amount of time from the start of the recovery process (usually measured in days), conditioned on the amount of the initial loss of functionality. System functionality is evaluated by considering the supply, demand and the electric power dispatch in the EPSS-Community system. Thus, by modeling the supply, demand and deliverable power, the electric power deficit can be tracked through the recovery phase, and the associated measures of EPSS-Community system resilience can be computed (Sun et al., 2015a, Didier et al., 2015).

Development of RFs is challenging. As opposed to VFs, there are significantly fewer RFs available in the literature (HAZUS, 2015). Furthermore, given the scarcity of post-earthquake recovery data, it is difficult to validate and calibrate the proposed RFs. Furthermore, generalizing RFs across EPSSs may not be possible because the component behavior is contingent on the unique characteristics of the EPSS they belong to. In addition, there is a remarkable influence of the state of the interdependent infrastructure systems, e.g. the transportation system, on the EPSS component and system repair activities during the post-earthquake recovery period. Most significantly, the interplay and coordination among different involving parties will govern the restoration of the damaged EPSS, and in turn, be reshaped by it.

In this paper, a novel approach to develop model-based RFs is proposed with the goal to dissect the complexities associated with the seismic recovery process of EPSS-Community system. This approach is based on the Agent-Based Model (ABM) paradigm, which has grown as a modeling strategy for capturing the dynamic behavior of a broad range of social, economic and ecological systems (O'Sullivan and Haklay, 2000). Specific agents representing the actors in the EPSS repair process, namely, the Operator of the EPSS and the Administrator of the community, are instantiated and the rules that govern the interaction among these agents are defined. The proposed framework is exemplified in a case study, where parameter variations are carried out to examine the influence of different agent behavior characteristics and the earthquake intensity on the EPSS supply recovery rate.

Using the ABM RFs and the compositional resilience quantification framework, the following questions can be answered:
1. How does the EPSS-Community system seismic recovery process evolve over time, and which parameters and decision influence it most?
2. How does the interaction among different agents involved in the recovery process of the EPSS-Community system shape that process?
3. What can be done to increase the resilience of an EPSS-Community system and thereby reduce the risks it is exposed to?

In this paper, Section 2 elaborates the principles of the proposed seismic resilience framework and the corresponding implementation. In Section 3, the framework is applied to an example EPSS-Community system. The topology, the parameter distributions, the electric power dispatch model as well as the systemic resilience measures are presented. Section 4 discusses the resulting seismic resilience of example EPSS-Community system under different earthquake scenarios and the interaction patterns between the agents. Section 5 summarizes the main findings and indicates possible future research topics.

## 2. FRAMEWORK FOR QUANTIFYING SEISMIC RESILIENCE OF EPSS-COMMUNITY

For an EPSS in normal operations, the electric power supply capacity is designed to cover the community demand. A strategy for distributing the generated electric power to the consumers with the goal of minimizing the costs, maximizing the profits, and minimizing the risks to the network infrastructure while minimizing the risks of disrupting the balance of supply and demand (i.e. brown- and black-outs) is called the electric power dispatch (Morsali et al., 2015). The balance between the supply and the demand can be disrupted due to damage from strong earthquakes. After such an event, the EPSS-Community system absorbs the seismic shock and enters a recovery phase until the system functionality stabilizes at a new level. The time scales of the absorption and the recovery phases are very different (hours vs. days), especially after catastrophic earthquakes (Ge et al., 2010, Iuchi et al., 2013) when the recovery phase may take years. Therefore, the seismic resilience of the EPSS-Community system is modeled by examining the functionality loss in the absorption phase as well as by tracking the functionality restoration path during the recovery phase separately for the demand and the supply sides of the EPSS-Community system.

The functionality of the EPSS-Community system is assessed by disaggregating this large-scale heterogeneous system into a set of supply and demand nodes while preserving their connectivity. The earthquake ground motion intensity measures (IMs) at the nodes differ and depend on their geographic location with respect to the earthquake epicenter. The geographic distribution of IMs at the EPSS-Community system node sites, the earthquake scenario, is determined using ground motion prediction equations for a given earthquake magnitude and location of its epicenter.

### 2.1. EPSS-Community system functionality assessment in the absorption phase

Given an earthquake scenario, seismic fragility functions are used to evaluate the Damage State (DS) of each component conditioned on the earthquake ground motion intensity the component experienced. Specific fragility functions are adopted for each component of EPSS (e.g. transformers, switches, and circuit breakers) and communities (e.g. office and apartment buildings, factories, schools, hospitals) to represent their unique physical and societal behavior (Déle and Didier, 2014). For simplicity, three damage states are considered, i.e. no damage (DS1), moderate damage (DS2) and extensive damage (DS3). Vulnerability of each component is then assessed by computing its remaining degree of functionality. In particular, no loss of functionality is assumed if a component is in damage state DS1, complete loss of functionality is assumed if the component is in DS3, and an intermediate interpolated degree of function loss is assumed for components in DS2. One such loss of functionality damage-state-based assignment is adopted in Sun et al. (2015a).

Decreased functionality of each node, on the supply or on the demand side, is determined individually, as specified above. Decreased functionality the EPSS-Community system at the end of the seismic damage absorption phase is computed by aggregating the functionality of the nodes using a model of EPSS operation, the seismic contingency electric power dispatch.

## 2.2. Recovery of the Community power demand

Once the earthquake damage is absorbed, the Community and the EPSS enter a relatively long recovery period. The post-earthquake recovery path of the electric power demand generated by the Community is affected by many factors, e.g. the efficiency of the reconstruction of buildings and industrial facilities, the restoration of public services, as well as by the societal norms and conventions. RFs are used to model the post-earthquake recovery of the community demand for electric power. RFs are the "mirror images" of the VFs in that they quantify the probability that a community component functionality, and thus its demand for electric power, will be restored after a certain recovery time, conditioned on the component damage state. As shown in Fig. 1, $RP_{DS2}(t)$ and $RP_{DS3}(t)$ RFs are sigmoidal functions of time in the recovery phase, formulated as lognormal probability distribution functions to satisfy the bounds and the monotonicity requirements. Following (Yang et al., 2012), a random number $r \in [0, 1]$ is generated at any point in time $t$ during the recovery simulation process. If a component is in DS2, it is considered as fully recovered at time $t$ if $r < RP_{DS2}(t)$. If a component is in DS3, it can recovery partially (to DS2) or fully (to DS1). Thus, the state of a component in DS3 is determined as follows: it recovers fully if $r < RP_{DS3}(t)$, it recovers partially, to DS2, if $RP_{DS3}(t) < r < RP_{DS2}(t)$, and it does not recovery if $r > RP_{DS2}(t)$.

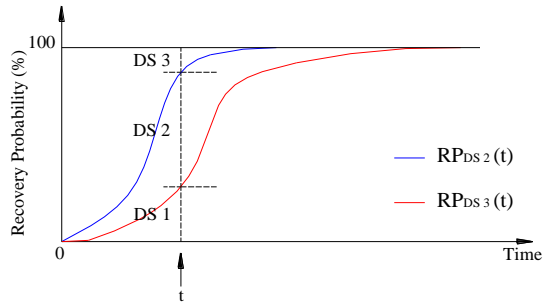

**Fig. 1. Recovery functions for the EPSS-Community system components.**

To facilitate quantification of the seismic resilience of EPSS-Community system, the power deficit for different sectors of the community is tracked separately. The community built inventory associated with the population through permanent or high intermittent occupancy, namely the residential buildings, the schools and the critical facilities such as hospitals, are combined into the *Population* sector, while the portion of the built inventory associated with production, such as industrial facilities and office buildings, are grouped into the *Factory* sector. The total demand at a distribution node $k$ is:

$$D(t)_k = D_p(t)_k + D_f(t)_k \tag{1}$$

where $D_p(t)_k$ and $D_f(t)_k$ are the instantaneous demands from the *Population* and *Factory* sectors connected to the EPSS at distribution node $k$, respectively, at time $t$ in the recovery process. Instantaneous demand depends on the initial damage and the rate of the recovery process of each component of the community built inventory, as well as on the occupancy type and quantity associated with that component. The initial damage state is determined as stated in Section 2.1, and the recovery process is modeled as stated above, sequentially in time throughout the post-earthquake recovery process. The electric power demand is computed as directly proportional to the component loss of functionality as proposed in Sun et al. (2015b).

The power delivered to the distribution node $k$ is determined by the electric power dispatch strategy adopted by the EPSS operator. In normal operating conditions, the delivered power $DP(t)_k$ matches the demand $D(t)_k$ at each distribution node $k$. Further, when a distribution node supplies both the *Population* and the *Factory* sectors of the community, the supply is distributed to the two sectors in proportion to their demand. In emergency situations, such as after an earthquake, the power deliverable to a distribution node may be smaller than the demand at that node because of loss of power generation substations and transformers due to earthquake-induced damage and possible inability to transmit the available power through the damaged EPSS network because of transmission line capacity limits. Consequentially, the power deficit at the distribution node k is:

$$PD(t)_k = D(t)_k - DP(t)_k \geq 0 \qquad (2)$$

If the distribution node *k* failed and is has not recovered at time *t* the power deficit is equal to the node demand, i.e. $PD(t)_k = D(t)_k$. The power deficit at node *k* is equal to zero when the delivered power equals the demand. For distribution nodes that supply both the *Population* and the *Factory* sectors of the community the power deficit is proportioned in the same proportion as the demand, namely:

$$PD_p(t)_k = PD(t)_k \cdot (D\_P(t)_k / D(t)_k) \qquad (3)$$
$$PD_f(t)_k = PD(t)_k \cdot (D\_F(t)_k / D(t)_k) \qquad (4)$$

At the level of the EPSS-Community system, the power deficit $PD(t)$, $PD_p(t)$ and $PD_f(t)$ are computed by summing up the corresponding distribution node power deficits:

$$PD(t) = \sum_k PD(t)_k \qquad (5)$$
$$PD_p(t) = \sum_k PD_p(t)_k \qquad (6)$$
$$PD_f(t) = \sum_k PD_f(t)_k \qquad (7)$$

Similarly, the system-level power demand *D(t)* and the deliverable power *DP(t)* are also obtained by aggregating the corresponding nodal values.

### 2.3. Lack of Resilience Measures

Resilience of the EPSS-Community system in the proposed framework is quantified by counting the number of people without power at time *t* during the post-disaster recovery process. This is an instantaneous resilience measure.

At the distribution node *k*, the number of people without power $PwoP(t)_k$ is directly related to the power deficit of the *Population* sector of the community. However, the actual number of people without power after a strong earthquake is contingent on many different factors (e.g. casualties, rescue and evacuation of the population, post-earthquake aid and temporary housing, long-term post-disaster organization of the community) that are not contained explicitly in the proposed framework. To estimate number of people directly affected by a power deficit at distribution node *k*, the following is assumed: a) if the transformers in the distribution substation failed, the entire population $P_k$ served by this node is considered affected; b) otherwise, the number of people without power is proportional to the ratio of the power deficit $PD(t)_k$ and the power demand $D(t)_k$ associated with the population. It is further assumed that 65% of the power demand is directly consumed by the residents (Eurostat, 2015) and the remaining 35% is consumed by other activities in the *Population* sector (e.g. transportation, food safety and preparation, heating or cooling, etc.). Therefore, for every operating distribution node *k* the percentage of population without power is:

$$PwoP(t)_k = (PD_p(t)_k / (0.65 \times D_p(t)_k)) * P_k \leq P_k \qquad (8)$$

where $P_k$ is the number of people served with distribution node *k*. This number corresponds to the number of the built inventory components served by the distribution node *k* and the occupancy of these buildings. It is assumed that $P_k$ remains constant during the entire recovery process, i.e. changes in the population (injuries, deaths, outflows and inflows due to evacuations, etc.) are not modeled.

At the EPSS-Community level, the number of people without power in the entire community is evaluated as *PPwoP(t)*, the percentage of population without power, calculated by summing the distribution node $PwoP(t)_k$ values at time *t* after an earthquake and normalizing by the total community population served by the EPSS. Another instantaneous system-level resilience measure is the electric power deficit, as shown in Equations 5 through 7. In addition, following the framework established by Ouyang et al. (2012), the gross functionality loss on the system level $F_{loss}(t)$ will also be employed as another direct, EPSS-related, measure of the lack of resilience:

$$F_{loss}(t) = 100 * (1 - G(t)/G_o) \qquad (9)$$

Here, $G_o$ and $G(t)$ refer to the level of functionality of the power generation capacity of the EPSS before and at a time *t* after the earthquake event, respectively.

## 2.4. Agent-based seismic recovery model of the EPSS supply

The post-earthquake recovery process of the EPSS supply is stochastic and case-specific. The information about the EPSS system state and component condition is assumed to reach the EPSS operator in the immediate aftermath of the earthquake. The operator assess situation and, after a short period of time, the repair teams are dispatched to repair EPSS components and restore power supply following a certain prioritization strategy. The rate of EPSS component repair depends on the functionality of other civil infrastructure systems, namely the telecommunication and the transportation systems, which is, in turn, affected by the lack of electric power, thereby inducing additional dynamics in the already complex EPSS-Community system.

For an EPSS in normal operations, the total electric power supply capacity is designed to cover the community demand, and the electric power dispatch is designed to minimize electricity costs and risk to the network infrastructure, as well as to maximize the EPSS profit. However, after strong earthquakes, it is likely that power generation and supply capacity cannot cover the demand from the served community. Therefore, the EPSS operator devises a Seismic Contingency Dispatch Strategy (SCDS) to distribute the available power resources to the consumers that can use electric power. Restrained by the inadequate power supply capacities, the operators first have to develop a "ranking list" (which can evolve over time) to decide which consumers should be prioritized, and which would be "sacrificed". This prioritization strategy is case-specific and it is established by making trade-off among societal, economic, political and sometimes even ethical considerations.

The EPSS recovery priorities and the SCDS might not necessarily reflect the needs of the community it serves. Therefore, the EPSS recovery process may need to be steered externally, by local community leaders, to ensure that community priorities are addressed (Opricovic and Tzeng, 2002, Kapucu and Liou, 2014). The actual recovery path results from the interplay between the EPSS and the community recovery priorities. The RF-based approach used model the recovery of the community demand is neither broad nor adaptable enough to model the recovery of the EPSS supply and the EPSS-Community system interactions during the recovery process. Thus, the post-earthquake restoration path of EPSS supply and the EPSS-Community system is modeled using an Agent-Based Model (ABM).

For simplicity and with no loss of generality, two principal players, the EPSS operator (hereafter Operator) and the local community leadership (hereinafter Administrator) are considered in this framework. The two-player framework can be easily extended to a multi-player one if the actions of additional entities that affect the recovery process are to be accounted for in the recovery process. The two featured agents act as follows:

**Operator:** The behavior of this agent is described by three attributes, namely Velocity, Efficiency and Tenacity, denoted as $V$, $E$ and $T_o$, respectively. Specifically, $V$ describes the average travelling speed of the repair team that is set to travel the shortest distance (a straight line) between two substations during the restoration campaign, $E$ quantifies the repair rate as the percentage of component functionality restored per day, and $T_o$ refers to the degree with which the Operator agent is capable of executing its own repair plan priorities.

**Administrator**: The behavior of this agent is described by one attribute, the Tenacity (denoted as $T_a$) that quantifies the ability of the community leaders to enforce community repair priorities at one (or more) community resilience measure threshold values.

Following an earthquake that disrupts the EPSS-Community system, the Operator agent starts the repair actions after an idle period needed for the emergency actions, EPSS state acquisition, and planning that includes the information about the state of other community infrastructure systems. The EPSS recovery plan reflects the balance of income and expenses deemed optimal from the business perspective of the EPSS owners. In this ABM, the tempo of the restoration is governed by the Operator agent's $V$ and $E$ parameter values that can be considered as directly proportional to the cost of recovery.

Simultaneously, the recovery of the community power demand proceeds (modeled as described in Section 2.2). However, the recovery priorities of the community may be different than those of the EPSS. For example, the need for electric power in the most damaged regions may be essential for emergency rescue, medical care, water and food supply, and for sanitation, making a prolonged lack of electricity supply in certain areas undesirable. The community may have its own recovery performance objectives (SPUR, 2009, Smith, 2013) that quantify the state of the built environment and civil infrastructure systems, as well as high-level functions of the community (Mieler et al., 2014), during the recovery process and set a recovery timeline and milestones.

A periodic check of the community recovery milestones is implemented in the proposed ABM through a comparison of one or more resilience measures to their threshold values at certain instances during the recovery process. If the rate of recovery is satisfactory, the EPSS operator is allowed to continue the recovery process following their own priorities. However, if the rate of recovery is too slow, the community may be able to enforce its recovery priorities by making the EPSS operator change its recovery plan. In the proposed ABM, this process is implemented through the interaction between the Administrator and the Operator agents. Namely, if the rate of community recovery is not fast enough (i.e. the resilience measure threshold values are not attained at the time they are checked), the Tenacity $T_a$ parameter of the Administrator is increased to make it more likely that the Operator will be incentivized to address the community recovery priorities. If this is the case, not only is the Operator recovery plan changed to address the community priorities, but the V and E parameters of the Operator agent are also incremented to model the increase in EPSS operator resources invested in community recovery.

In the ABM framework proposed in this paper, the recovery of the EPSS-Community system is evaluated using the percentage of people without power *PPwoP(t)* resilience measure described in Section 2.3. The rate of recovery is evaluated only once, at *t*=72 hours (three days) after the earthquake disaster (SPUR, 2009, Smith, 2013). This point in the recovery process is assumed to be critical for the success of the recovery process and is termed the *Resilience Check Time* (Cimellaro et al., 2016). However, other resilience measures can be checked at one or more other instances during the recovery process to make sure that the recovery process is meeting the community performance objectives.

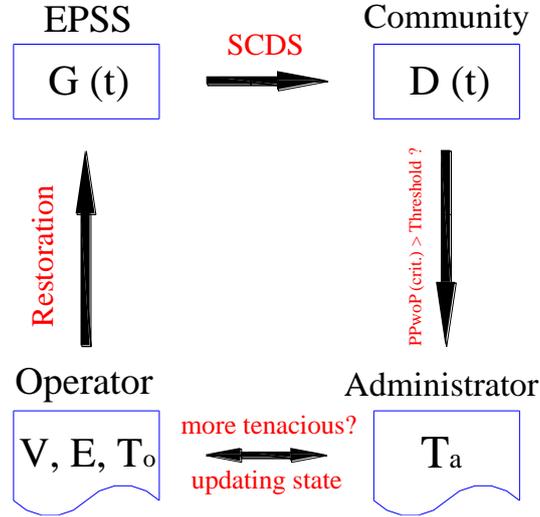

**Fig. 2. Illustration of the interactions among Administrator and Operator agents in the proposed ABM.**

The value of *PPwoP(72 hours)* is compared to the community recovery measure threshold value to check the recovery progress. If the progress is satisfactory, the Operator agent continues with the recovery process following its original repair priority plan. Otherwise, the Tenacity parameter of the Administrator agent is incremented relative to the initial Tenacity of the Operator. The values of the Tenacity attributes of the two agents determine whose priorities are going to be addressed first during the recovery process, as shown in Fig. 2. Namely, if $T_a < T_o$ there are no changes, but if $T_a \geq T_o$ the priorities of the community take precedence. As a result, it is likely that the Operator will be incentivized to embrace the Administrator's recovery priorities. For example, the Operator may need to repair the most seriously damaged substations first, and only thereafter proceed in the descending order of damage for the remaining damaged EPSS components. If the repair plan change is triggered, the speed *V* and the efficiency *E* attributes of the Operator are increased in order to increase the rate of the recovery process. The agents' state update also reflects the ability of the Administrator to prioritize the EPSS Operator vehicles on the available roads while coordinating the restoration campaign. The EPSS restoration proceeds using the plan selected after the check of the recovery process progress until all damaged EPSS components are fully repaired.

## 3. APPLICATION OF THE PROPOSED ABM FRAMEWORK

The proposed ABM framework was implemented in *Matlab* and is demonstrated using a virtual EPSS-Community system. The EPSS is extracted from the IEEE 118-node Benchmark System (Christie, 1993). As illustrated in Fig. 3, the EPSS consists of 15 generation substations (red squares) and 19 distribution substations (blue circles). Generation substations inject electric power into the network and transfer the electric power using the interconnecting power lines. Distribution substations supply the low-voltage power grids, i.e. extract electric power from the network and transfer it to the consumers in the community. The nominal electric power supply capacity of the 15 generation substations and the electric power demand of the 19 distribution substations are presented in Table I, respectively. Also listed is the assumed population served by each distribution substation. For simplicity, only the distribution and the generation substations are considered, while the other components of the EPSS as not modeled. The topology and the societal structure of modern communities are complex and heterogeneous. In order to reduce the complexity of this example, the community served by the virtual EPSS described above is disaggregated into two sectors, i.e. the Population sector comprising the residential buildings and the critical facility, and the Factory section, comprising the industrial facilities.

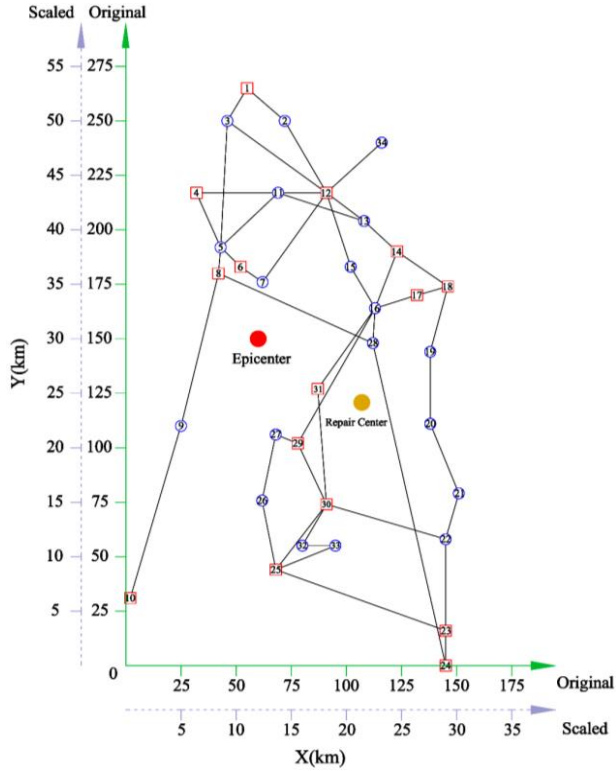

**Fig. 3. Topology of the virtual EPSS-Community system.**

The built inventory in each of these sectors is further divided into sub-categories listed in Table II. The EPSS and the served community are made geographically denser by scaling the IEEE 118 EPSS down by a factor of 5, resulting in a roughly 32x52km area of the virtual EPSS-Community system (Fig. 3). The scaling ensures that significant ground shaking from earthquakes with magnitudes between 4 and 7 affects most of the EPSS and of community component locations in most simulations, thus generating non-trivial simulation outcomes.

**Table I. Characteristics of the generation and distribution substations of the virtual EPSS.**

| Generation | | Distribution | | | |
|---|---|---|---|---|---|
| Generation Substation | Power Generation (MW) | Distribution Substation | Power Demand (MW) | Population Served $P_k$ ($10^3$) | Sector Served |
| 1 | 150 | 2 | 41.6162 | 76 | *Population+Factory* |
| 4 | 50 | 3 | 61.7809 | 66 | *Population+Factory* |
| 6 | 50 | 5 | 69.6968 | 96 | *Population+Factory* |
| 8 | 100 | 7 | 16.1609 | 66 | *Population* |
| 10 | 50 | 9 | 52.9856 | 32 | *Population+Factory* |
| 12 | 100 | 11 | 34.3699 | 134 | *Population* |
| 14 | 25 | 13 | 53.3392 | 224 | *Population* |
| 17 | 50 | 15 | 60.0792 | 224 | *Population* |
| 18 | 50 | 16 | 64.6848 | 256 | *Population* |
| 23 | 25 | 19 | 4.1300 | 24 | *Population* |
| 24 | 50 | 20 | 26.9800 | 24 | *Population+Factory* |
| 25 | 50 | 21 | 6.1100 | 24 | *Population* |
| 29 | 50 | 22 | 4.1300 | 24 | *Population* |
| 30 | 50 | 26 | 29.2884 | 76 | *Population+Factory* |
| 31 | 50 | 27 | 51.3934 | 72 | *Population+Factory* |
| | | 28 | 16.6912 | 64 | *Population* |
| | | 32 | 16.0809 | 66 | *Population* |
| | | 33 | 9.4406 | 44 | *Population* |
| | | 34 | 114.1500 | 0 | *Factory* |

**Table II. Community built inventory types in the two power demand sectors.**

| Sector | Use | Type of Structure |
|---|---|---|
| *Population* | Residential | Reinforced concrete apartment building |
| | | Masonry apartment building |
| | | Masonry single-family house |
| | Critical Facility | Hospital |
| | | School |
| *Factory* | Industrial | Heavy industrial building |
| | | Light industrial building |
| | | Office building |

The seismic hazard environment is modeled by locating the earthquake hypocenter close to the geographic center of the EPSS system, as shown in Fig. 3. This hypocenter location is not changed in this study to control the computational effort. The intensity of shaking at each EPSS or community component site, measured using peak ground motion displacement, velocity and acceleration values is computed using the ground motion attenuation relations proposed by Campbell and Bozorgnia (2008). The magnitude of the earthquake (*M*) is associated to the occurrence probability using the bounded Gutenberg-Richter law (the seismic hazard curve) with parameters $a$=4.4 and $b$=1 (Kramer, 1996).

### 3.1. Vulnerability analysis

Libraries of seismic fragility functions are available for the components of the community built environment (Rossetto and Elnashai, 2003, Kwon and Elnashai, 2006, Jeong and Elnashai, 2007, Senel and Kayhan, 2009, ATC-58, 2015, OpenQuake, 2015, Syner-G, 2015) and civil infrastructure systems (HAZUS, 2015, Syner-G, 2015) including the EPSS components. Fragility functions suitable for the built inventory (Table II) and the components of the EPSS generation and distribution substations (Table I), such the transformers, circuit-breakers and busses, are selected from these documents. The VFs are obtained as described in Section 2.1. For simplicity but with loss of generality, only DS1 and DS3 are considered for the EPSS components with the exception of the generators. Three damage states were used to describe the earthquake damage to the built inventory components and to evaluate the resulting electric power demand as described in Section 2.1. The power demand of schools and hospitals in DS1 and DS2 was kept at the pre-disaster level regardless of the incurred damage to reflect their role as emergent shelters.

### 3.2. Parameters of the power demand Recovery Functions

Power demand stems from the community requirements, i.e. from the electric power needed to support the functions of the components of the community built inventory (Table II). Therefore, the power demand depends on the ability to restore the functions within the buildings, which, in turn, depends on the level of the incurred earthquake damage. The RFs are lognormal probability distribution functions with the parameters defined in Table III. Different RFs are defined for damage states DS2 and DS3. It is assumed that no recovery is needed if the component is in damage state DS1. Given the component damage state, the probability of its full function recovery at time *t* in the recovery process is computed using the RF appropriate to the specific building inventory. The resulting electric power demand is computed as outlined in Section 2.2.

**Table III. Parameters for the RFs of the community built inventory components.**

| Type of Structure | Damage State | Mean (Days) | Std. (Days) |
|---|---|---|---|
| RC apartment building | DS2 | 14 | 12 |
| | DS3 | 210 | 60 |
| Masonry apartment building | DS2 | 14 | 12 |
| | DS3 | 210 | 60 |
| Masonry single-family house | DS2 | 10 | 9 |
| | DS3 | 150 | 54 |
| Hospital | DS2 | 12 | 10 |
| | DS3 | 150 | 30 |
| School | DS2 | 25 | 20 |
| | DS3 | 240 | 90 |
| Light industry building | DS2 | 45 | 40 |
| | DS3 | 270 | 180 |
| Heavy industry building | DS2 | 45 | 40 |
| | DS3 | 300 | 180 |
| Office building | DS2 | 25 | 20 |

| | DS3 | 240 | 72 |

The recovery of the residential building and hospital functions is assumed to be the fastest among all community built environment components. Assuming that the population is not evacuated, the focus of the recovery is restoring shelter. Once the people are in their homes, or temporary shelters, their demand for electricity will recover quickly. The recovery of multi-story apartment buildings is assumed to take more time than the recovery of single-family homes. The durations of the recovery of light and heavy industry is the longest but is also affected by a larger uncertainty, because their operation cannot fully recover, or even restart, unless the community population (not only EPSS) is on its way to recovery. The schools and high-rise buildings can recover moderately fast, compared to the other built environment components mentioned above. The recovered power demand $D(t)$ at the EPSS-Community system level is tracked by integrating the recovered demand of every component of the community built inventory as described in Section 2.2.

### 3.3. Parameters of ABM agents

The parameters of the ABM Operator and Administrator agents are defined as random variables with probability distributions shown in Table IV. The velocity of the Operator agent repair crew vehicle $V_i$ is scaled down by the length scale factor to match the size of the scaled-down EPSS-Community system (Fig. 3), and it also accounts for the damage to the road network that may hamper the repair efforts. The initial repair efficiency $E_i$ quantifies the initial recovery efficiency of the generation substations. To simplify the simulations, the recovery of the distribution substations was modeled in parallel by assuming that all of them will fully restore their functionalities at a specific period of time, the Recovery Threshold, after the earthquake event because of their proximity to the inhabited areas and the availability of additional repair crews. The Recovery Threshold for distribution substations is defined by a uniform distribution with limits set as a function of the earthquake magnitude (Table IV).

**Table IV. Probability distributions of the Operator and Administrator agent parameter values.**

| Parameter | Distribution Type | Lower Limit | Upper Limit |
| --- | --- | --- | --- |
| **Operator** | | | |
| $V_i$ (km/h) | Uniform | 6 | 8 |
| $E_i$ (1/day) | Uniform | 20% | 30% |
| Recovery Threshold (day) | Uniform | $M$-2 | $M$+4 |
| $T_{o,i}$ | Uniform | 0.4 | 0.5 |
| **Administrator** | | | |
| $T_{a,i}$ | Uniform | 0.3 | 0.4 |

\* $M$ is the seismic magnitude obtained from the hazard curve.

### 3.4. Monte Carlo simulation of the recovery of the EPSS-Community system

The uncertainties related to the earthquake scenarios, the EPSS and the community component damage states, the disaster preparedness of the EPSS operator and the community, and the interaction between the recovery governance structures are represented by probability distributions and propagated via Monte Carlo simulations of individual earthquake damage recovery scenario realizations.

Each simulation starts by assuming an earthquake of a certain magnitude $M$ occurred with the hypocenter located as shown in Fig. 3. The probability of occurrence of this earthquake is obtained from the seismic hazard curve. The ground motion intensities are computed using the Campbell and Bozorgnia (2008) attenuation relations for the rock site. The VFs (Section 3.1) are used to establish the state of the EPSS-Community system at the beginning of the recovery phase. This state description comprises of the damage state of each EPSS and community component. Before the recovery time $t$ counter is initiated, the SCDS of the EPSS Operator is selected (Table V) and the attributes of the Operator and the Administrator agents $V_i$, $E_i$, $T_{o,i}$ and $T_{a,i}$ are randomly generated from the distributions as detailed in Section 3.3. The Operator's initial tenacity $T_{o,i}$ is always larger than the Administrator's initial tenacity $T_{a,i}$ allowing the Operator to plan the EPSS restoration without interference. As a result, the repair crew will focus on the least seriously damaged substation, and then travels to fix the other substations following the ascending damage severity order. Such prioritization is justified on two grounds: 1) repairing slightly damaged components tends to be fast and the access to them is, most likely, not hampered by the damage to the transportation systems, enabling prompt restoration of electric power to some customers; and 2) heavily damaged substations may need replacement rather than repair, which requires additional planning, design, financing, equipment acquisition, and transport.

The damage severity $S_n$ of a damaged generation substation $n$ is quantified as the functionality loss normalized by the original functionality level, i.e.:

$$S_n = 100*(1-F_{d,n}/F_{o,n}) \tag{10}$$

where $F_{o,n}$ and $F_{d,n}$ are the levels of functionality (the power generation capacity) before and after the earthquake, respectively. The repair time $r_n$ for each generation substation $n$ is:

$$r_n = S_n/E_i \tag{11}$$

Given the repair priority order, the travel time $t_n$ between the damaged substations $n$-1 and $n$ for $n \in [2, N]$, can be calculated based on the road distance $D_n$ between substation $n$-1 and $n$ as $t_n = D_n/V_i$. For any damaged substation $n$ in the repair list, the traveling time to it and the time to repair it is:

$$\begin{cases} t_1 = t_0 + d_0/V_i \\ t_n = r_{n-1} + D_n/V_i & \text{if } T_{a,i} < T_{o,i} \\ r_n = t_n + S_n/E_i & \text{if } T_{a,i} < T_{o,i} \\ t_n = r_{n-1} + D_n/V & \text{if } T_a \geq T_{o,i} \\ r_n = t_n + S_n/E & \text{if } T_a \geq T_{o,i} \end{cases} \tag{12}$$

where $t_0$ is the emergency action and restoration planning period right after the earthquake (set to equal 36 hours in this study) and $d_0$ is the distance between the repair center and the first-generation substation to be restored. The attributes $V_i$ and $E_i$ in Eq. (8) are updated if the state of the Operator is updated after the recovery performance check at the *Resilience Check Time* as described in Section 2.4. Specifically, the Tenacity parameter $T_{a,i}$ of the Administrator agent is incremented by 0.1, i.e. $T_a = T_{a,i} + 0.1$, at the *Resilience Check Time*, and compared to the Tenacity parameter $T_{o,i}$ of the Operator agent. The magnitude of the increment value is selected considering the values of the Tenacity Parameter distribution bounds in Table IV. If the community concerns prevail $T_a \geq T_{o,i}$ the EPSS component repair priority list is reversed (i.e. the most damaged substation will be repair first), and the Velocity and the Efficiency parameters of the Operator agent are increased (in this study $V = 1.1*V_i$ and $E = 2*E_i$) to increase the rate of the recovery process.

The seismic restoration campaign evolves though the discrete time steps $t_n$ and $r_n$. The generation capacity $G(t)$ at the level of the EPSS-Community system at time $t$ is:

$$G(t) = \begin{cases} \sum_{j=1}^{n-1} f_{o,j} + f_{d,n} + \frac{t-t_n}{r_n-t_n} \times (f_{o,n} - f_{d,n}) & t_n \leq t < r_n \\ \\ \sum_{j=1}^{n-1} f_{o,j} + f_{d,n} & r_{n-1} \leq t < t_n \end{cases} \tag{13}$$

During time step $t_n$, power supply capacity $G(t)$ remains unchanged while the repair crew is travelling because no substation is under repair. Once the repair crew arrives at the generation substation $n$ and starts the repair process, its functionality follows a linear recovery with slope equal to $E_{ini}$, if $t_n \leq t < r_n$ and $T_a < T_{o, ini}$, or with slope equal to $E$, if $t_n \leq t < r_n$ and $T_a \geq T_{o, ini}$.

The transmission lines are considered to remain undamaged in this model. The distribution substations are considered to be fully functional after the Recovery Threshold time (Table IV). Therefore, the power demand at the level of the EPSS-Community system $D(t)$ is computed as stated in Section 3.2.

### 3.5. Seismic Contingency Dispatch Strategy

In this case study, five different SCDSs listed in Table V are proposed. Strategy 1 prioritizes the supply to communities which have the largest post-earthquake demand. This strategy reflects the EPSS operator preference to supply first their largest users during the restoration process. Such strategy assumes that electric power can be produced in sufficient quantities.

**Table V. Prioritization Strategies for Seismic Contingency Dispatch (SDCSs)**

| Strategy | Criterion |
|---|---|
| 1 | Maximum demand |
| 2 | Minimum demand |
| 3 | Largest Population |
| 4 | Maximum normalized demand |
| 5 | Minimum normalized demand |

On the other hand, distribution substations with small power demand might be prioritized because the ability to generate electric power may be limited after an earthquake (SCDS 2). Similarly, SCDSs 4 and 5 prioritize distribution substations according to the maximum and minimum power demand normalized by the pre-earthquake demand, respectively. Finally, Strategy 3 prioritizes distribution substations serving large populations.

In order to supply the communities following the priority list established by the SCDS, the distribution substation nodes are ranked in the descending order according to the prioritization criterion, e.g. the instantaneous power demand for SCDS 1. The capability of the EPSS to transmit power to distribution substation $k=1, 2,…, 19$ after the disruption, is assessed by evaluating the shortest paths between a distribution substation $k$ and any generation substation $l=1, 2,…, 15$. In order to allocate the available transferable power, the demand of the distribution substation is assigned to the closest generation substation following the dispatch priority ranking. The power allocation terminates if the total demand of the distribution substations is satisfied, or if the EPSS generation capacity is reached. In the latter case, distribution substation $k$ cannot be supplied and is affected by an electric power deficit $PD(t)_k$. The power allocation procedure and the available transferable power are updated during the recovery process based on the updated status of supply and demand side EPSS and community components.

## 4. RESULTS OF THE VIRTUAL EPSS-COMMUNITY SYSTEM RESILIENCE ASSESSMENT

The recovery of the virtual EPSS-Community system is simulated using the Monte Carlo simulation technique. Two case studies are investigated. In Section 4.1, the Operator agent guides system recovery and its repair priorities are always enforced. The influence of the five SCDSs on the EPSS-Community system resilience is assessed in Section 4.2. In Section 4.3, the Administrator agent is introduced and its influence on the recovery path is assessed by the comparison with the results of Section 4.1. This comparison is done only for SCDS 1. Each case study involves 2000 earthquake events where damage is induced and the resulting recovery paths are tracked in time until pre-earthquake conditions are restored. The results of these 2000 MC simulations are then statistically aggregated.

### 4.1. Operator Agent Recovery Process

Fig. 4 shows the medians of the electric power generation capacity, the delivered electric power, the electric power demand and the electric power deficit in a scenario with an earthquakes of magnitude $M=7.5$ and SCDS 1.

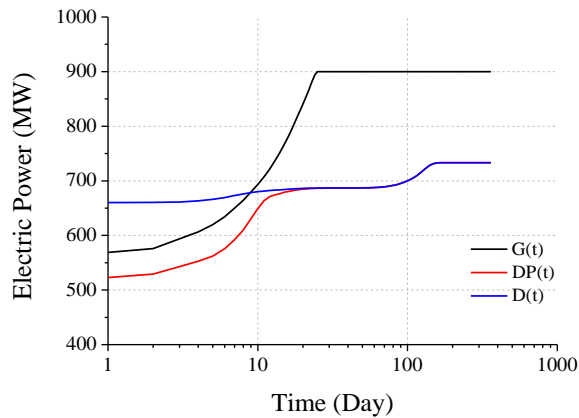

**Fig. 4. Evolution of the generation capacity (*G(t)*), deliverable power (*DP(t)*), and power demand (*D(t)*) in the EPSS-Community system for a Magnitude 7.5 earthquake scenario and SCDS 1 (median values).**

Before the event, the EPSS supplies 900 MW of the electric power and covers the 733 MW of community demand. Immediately after the event, the median electric power demand drops to 660 MW and the median power generation capacity drops to 569 MW. Further, the median deliverable power drops to 524 WM due to low-voltage power transformer failures and high-voltage power transmission line capacity insufficiencies. After the recovery starts, it takes 25 days to restore the median EPSS generation capacity to the pre-earthquake level, whereas it takes 165 days for the community demand to recover.

The evolution of the *PPwoP* EPSS-Community system resilience measure is plotted in Fig. 5 for the *M*=7.5 earthquake scenario and SCDS 1. Immediately after the strong earthquake, 34% of the community population is likely to be without electric power, as indicated by the median *PPwoP*. The median duration to complete recovery, i.e. the time required for all people in the community that can use electric power to receive it, is 11 days. The 20% and 80% quantile curves in Fig. 5 indicate the uncertainty associated with the seismic recovery of the EPPS-Community systems, and show that the median *PPwoP* is symmetric with respect to the quantiles.

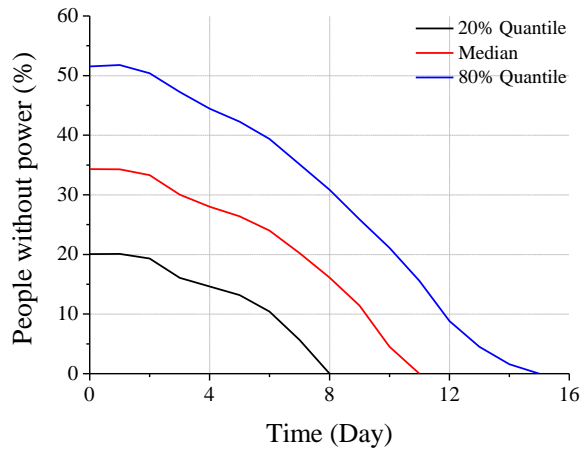

**Fig. 5. Evolution of PPwoP for a magnitude 7.5 earthquake scenario and SCDS 1.**

The influence of the earthquake magnitude on the EPSS-Community system resilience is assessed in Fig. 6, which shows the *PPwoP* for *M*=6, 6.5, 7 and 7.5 earthquake scenarios with SCDS 1.

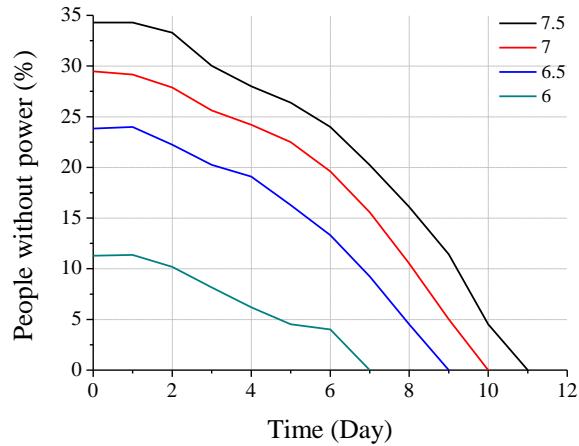

**Fig. 6. The evolution of the median PPwoP for different earthquake magnitude scenarios for SCDS 1 (median values).**

The resilience of the EPSS-Community system is significantly affected by the earthquake magnitude. The post-earthquake median *PPwoP* values are 34%, 29%, 24% and 11% for magnitudes 7.5, 7, 6.5 and 6, respectively. The median duration to complete recovery is 7, 9, 10 and 11 days for the four seismic scenarios, respectively. Fig. 6 reveals a remarkable gap between the consequences of *M*=6 earthquake and the three stronger earthquakes, indicating that robustness, i.e. ability to reduce the initial damage, is effective in increasing the resilience of the EPSS-Community system, but only up to the seismic hazard level to which the system components were designed for. Designing for stronger earthquakes is initially costly: such investment should be compared to the probable loss estimates and a cost-benefit analysis could be used to make a rational decision about an appropriate seismic design hazard level.

Fig. 6 shows that the electric power restoration process can be roughly divided into three stages after the EPSS repair starts (36 hours after the earthquake in this study). Specifically, in the first stage (between 36 and 72 hours after the earthquake) the *PPwoP* decreases quickly although the EPSS restoration strategy targets the least seriously damaged stations first. The explanation can be found if the demand-side is taken into account: the recovery probability for all the component of the built environment is rather small before the third day (see Table II) and therefore the gap between the deliverable and the demanded power can narrow relatively fast. In the second recovery stage between day 3 and day 5 (or day 6 for the *M*=6 earthquake scenario), the *PPwoP* keeps decreasing, but at a smaller rate, because buildings are restored and the power demand increases. In the third stage after day 5 (or day 6 for the *M*=6 earthquake scenario), the generation capacity is restored quickly because the repair crew focuses on more severely damaged generation substations, and the *PPwoP* decreases at a faster rate again.

### 4.2. Impact of the Seismic Contingency Dispatch Strategies (SCDSs)

The resilience of different community sectors, the *Population* and the *Factory* sectors specified in Table I, is affected by the SCDS implemented by the EPSS operator. In order to quantitatively examine this influence, the seismic resilience of EPSS-Community system is evaluated for the two sectors separately for the five SCDSs listed in Table V.

The main function of the *Population* sector is to support the basic livelihood of community dwellers. Its seismic resilience is measured by the *PPwoP* resilience measure defined in Section 2.3. The evolution the *PPwoP* median during the recovery process in the *M*=7.5 earthquake scenario is presented in Fig. 7 for the five SCDSs. The five curves in Fig. 7 form two clusters. The first cluster results from SCDSs 1 and 4, which prioritize the power supply to communities having large demand and large normalized demand. They result in *PPwoP* of 34% and 38% immediately after the earthquake, respectively. It takes 11 days to fully restore the power to the *Population* sector under these two strategies. Conversely, SCDSs 2, 3 and 5 lead to much smaller *PPwoP* (less than 20%) immediately after the earthquake, and a faster recovery of the *Population* sector. Therefore, the seismic resilience of the *Population* sector is improving significantly faster if SCDSs 2, 3 (and to some extent 5) are employed by the EPSS operator.

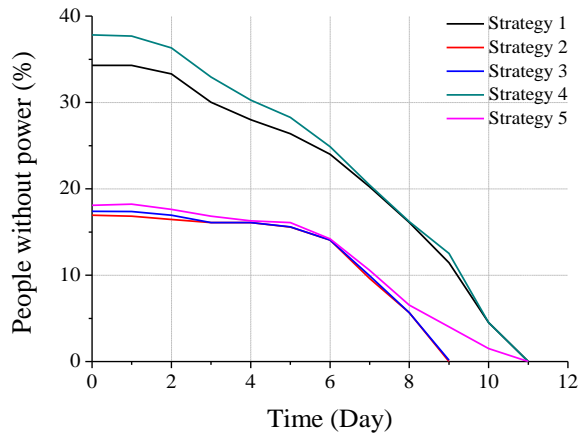

**Fig. 7. Evolution of the *Population* sector recovery for the M=7.5 earthquake scenario and the five SCDSs (median values).**

On the other hand, the choice of the SCDSs has an opposite effect on the recovery of the *Factory* sector. This recovery process is tracked using the evolution of the power deficit of the *Factory* sector, $PD_f(t)$ defined in Eq. (7), shown in Fig. 8, for the five SCDSs. In particular, the *Factory* sector power deficit after the earthquake is 19 MW for SCDS 1 and 22 MW for SCDS 4. The SCDSs 2, 3 and 5 result in over 75 MW power deficit of the *Factory* sector. Furthermore, using SCDS 1 fully satisfies the demand of the *Factory* sector in just 5 days, compared to between 9 and 10 needed when using other strategies.

Of the five consider strategies, SCDS 5 offers a balance between relatively high resilience and speedy recovery of the *Population* sector and somewhat shorter recovery of the *Factory* sector. This result can be explained in terms of the interplay among the earthquake scenario, the topology of EPSS-Community system, and the recovery pattern of different community sectors. First, the distribution stations with large power demand, e.g. node 34 in Table I, which is a purely *Factory* sector distribution node, and nodes 13, 15 and 16, which are mixed, will be prioritized in SCDS 1 and SCDS 4. Distribution node 34 is much farther away from the epicenter compared to distribution nodes 13, 15 or 16, making the expected damage to node 34 much smaller. Therefore, it is likely to recover significantly faster.

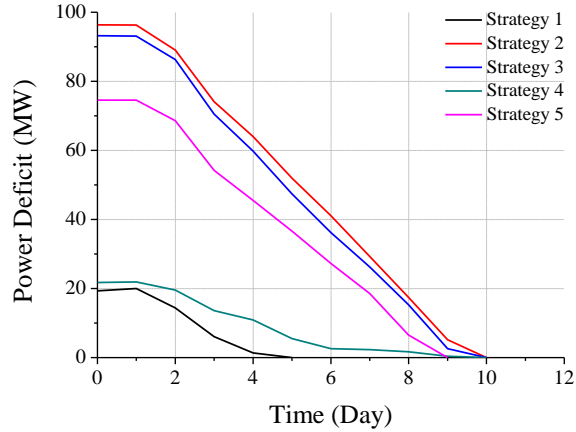

**Fig. 8. Evolution of the *Factory* sector recovery for the M=7.5 earthquake scenario and the five SCDSs (median values).**

### 4.3. Operator and Administrator Agent Recovery Process

The effects of the interaction between the Operator and the Administrator agents on the recovery of the EPSS-Community system are examined by re-running the magnitude *M*=7.5 earthquake SCDS 1 scenario simulations with both agents. As explained in Sections 2.4 and 3.4, a check of the rate of EPSS-Community system recovery is performed at the *Resilience Check Time* set to be 72 hours after the earthquake in this study. Then, the value of the *PPwoP* EPSS-Community system resilience measure is compared to a threshold. Three threshold values, 10%, 20% and 30%, are used in this study. If the *PPwoP* exceeds the threshold at the *Resilience Check Time*, the Tenacity Parameter of the Administrator agent is incremented and compared to the Tenacity Parameter value of the Operator agent, as outlined in Section 3.4. In the conducted simulations, the probabilities that the recovery plan of the Operator agent is changed were 46.8%, 38.6% and 26.0% for the 10%, 20% and 30% *PPwoP* threshold, respectively. For comparison, the case without activation threshold, i.e. no Administrator agent, is also reported.

Fig. 9 shows that the involvement of the Administrator agent increases the rate of power generation recovery and shortens the recovery process. In the case of a strict *PPwoP* threshold equal to 10%, the change of repair priorities induced by the Administrator agent resulted in complete recovery after 19 days as compared to 25 days that it takes without the Administrator agent. The effects of the Administrator agent on the recovered generation capacity decrease for larger *PPwoP* threshold, and the recovery path approaches the case without interaction as the Administrator becomes "weaker".

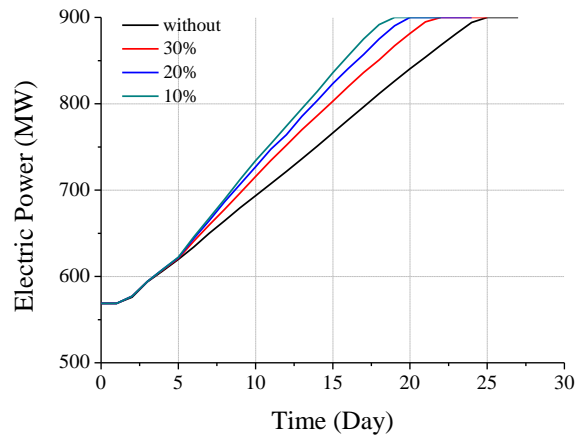

**Fig. 9. Evolution of the electric power generation recovery for the M=7.5 earthquake SCDS 1 scenario (median values).**

The effect of the Administrator agent is also evident in the rates of decrease of *PPwoP* during the recovery process, as shown in Fig. 10 for the *M*=7.5 earthquake SCDS 1 scenario. The change in the recovery rate does not occur immediately at the *Recovery Check Time* (3 days in this study), but somewhat later, on day 5. This change then results in shortening of the time to full recovery from 11 to 10 days, given the presence of the coordination of Administrator agent. Furthermore, like the power generation capacity demonstrated in Fig. 9, *PPwoP* was also found to decrease quicker when the Administrator became "stricter".

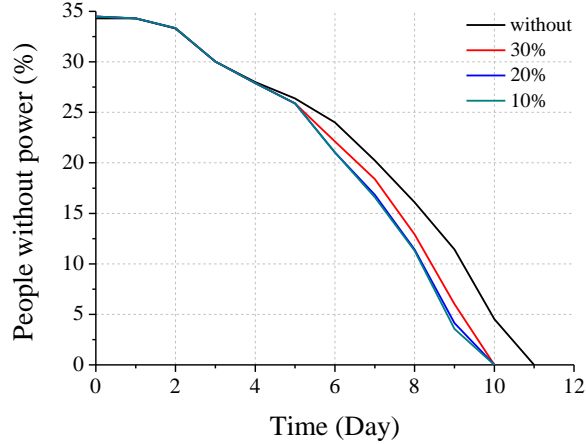

**Fig. 10. Evolution of the *PPwoP* EPSS-Community resilience measure for the M=7.5 earthquake SCDS 1 scenario (median values).**

### 4.4. Sensitivity analysis

The results of Section 4.1 and 4.3 show that the resilience behavior of the EPSS-Community system is heavily influenced by the magnitude of the earthquake scenario and by the resilience measure threshold of the Administrator agent. Thus, the sensitivity analysis is focused on these two parameters. To investigate the sensitivity of the systemic response to those parameters, the behavior of the system under different set of magnitude and threshold parameters is examined. Without loss of generality, the normalized systemic functionality loss $F_{loss}$ twenty days (Equation 9) and the *PPwoP* (defined in Section 2.3) ten days after the earthquake event, respectively, are presented in Fig. 11 for the earthquake M=7.5 SCDS 1 scenario.

Both measures are equal to zero (or very close to zero), if the magnitude is lower than or equal to 6.25. Therefore, they show no sensitivity to the Administrator agent threshold parameter in case of low-magnitude earthquake scenarios. However, for scenarios with earthquake magnitudes larger than 6.25, the functionality loss and *PPwoP* increase significantly, indicating that the impact of the threshold parameter is more pronounced. The functionality loss is zero and the *PPwoP* is about 3% when the *PPwoP* Administrator agent threshold is equal to 10%, even though the magnitude of the seismic scenario is 7.5 (this is also shown in Fig. 9). However, the resilience of the system decreases as the Administrator agent becomes less strict. For the *PPwoP* threshold value of 70%, the functionality loss reaches 8% and the *PPwoP* reaches 12%, and remains virtually constant for threshold values higher than 70%.

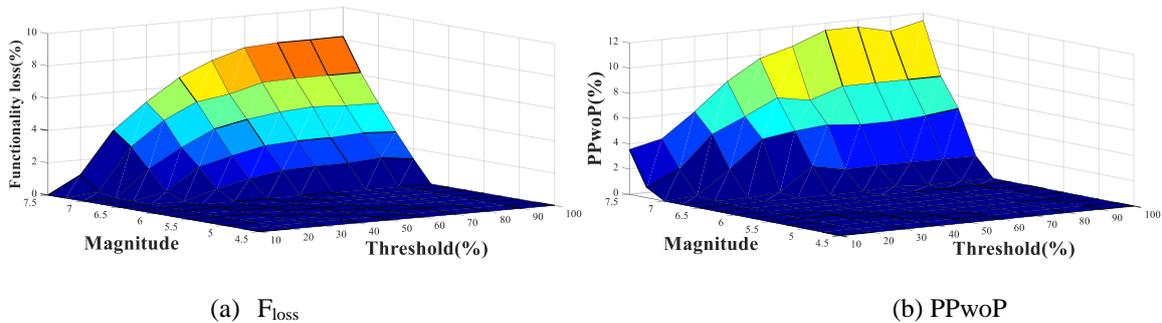

(a) $F_{loss}$          (b) PPwoP

**Fig. 11. Resilience behavior in terms of the median normalized functionality loss $F_{loss}$ twenty days (left panel) and the median *PPwoP* ten days (right panel) after the earthquake event for the EPSS-Community**

**system in earthquake scenarios with different magnitudes and Administrator agent *PPwoP* threshold values.**

## 5. CONCLUSIONS AND OUTLOOK

In this paper, an agent-based framework developed to quantify the seismic resilience of an Electric Power Supply System (EPSS) and the community it serves was presented. In this framework, the loss and the restoration of the EPSS electric power supply capacity and the electric power demand from the served community are used to assess the instantaneous electric power deficit at any point in time during the damage absorption and recovery process. Therefore, such framework enables the investigation on the risk exposure and resilience of a Community-EPSS system.

In this framework, damage to the components of the EPSS and of the community built environment is evaluated using the seismic fragility functions. The restoration of the community electric power demand is evaluated using the seismic recovery functions. The novelty presented in this paper is in the use of an agent-based approach to model the post-earthquake EPSS power supply recovery, namely the recovery of the generation and delivery capacity of the system. Namely, two agents, the EPSS Operator and the Community Administrator, are used to represent the recovery priorities, planning and actions of the EPSS operators and the community as they strive to recover from a disaster. The agent-based framework enables nuanced modeling of the recovery process, emphasizes the need to consider the supply and the demand occurring in the EPSS-Community system to understand the recovery process, and highlights the effects of EPSS system repair and electric power dispatch strategies and community demand recovery on the resilience of different sectors of the community and the community as a whole. Remarkably, the agent-based model reveals the emergence of possibly conflicting interests of the community and the EPSS operator. Furthermore, it is demonstrated that resolution of such conflicts would profoundly influence the recovery process of the EPSS-Community system. The communities could enforce post-earthquake recovery performance objectives by intervening, if the recovery is slow, through changing the recovery priorities of the CI operators, and thus actively reducing the community risk exposure (e.g. reduce the risk of a long-lasting power outage that affects a significant number of people).

In particular, the proposed agent-based seismic resilience quantification framework provides the following answers to the questions posed in the introductory section of this paper. The EPSS-system-related electric power deficit and the community-related resilience metrics are computable and can be tracked during the damage absorption and recovery processes to indicate the resilience (or lack thereof) of the EPSS-Community system. Clearly, the larger the earthquake hazard, the more challenged the EPSS-Community system will be. However, increasing the robustness to reduce the vulnerability of the EPSS and Community components, as well as increasing the rate of their repair results in an EPSS-Community system that has a smaller risk exposure and recovers faster. Remarkably, the rate of electric power demand recovery is at least as important for the EPSS-Community system resilience as the rate of electric power supply recovery. The community can monitor the rate of the EPSS-Community recovery process using the proposed resilience measures and intervene effectively early on in the recovery process to change its rate or priorities. The presented sensitivity analysis shows that the resilience of the system could be significantly improved if the Administrator agent steps in to emphasize the community priorities and demand speedier and more efficient repairs. This is particularly true when it comes to strong earthquakes.

The conducted simulations also reveal the remarkable role contingency electric power dispatch strategies have on the resilience of different community sectors. This points to the need to plan a dynamic post-disaster interaction between the community and the civil infrastructure system operators and develop contingency service dispatch strategies as well as consistent contingency demand regulation measures to shape the recovery process and maximize community resilience. The proposed agent-based seismic resilience quantification framework can be used to model and test such strategies and measures and, thus, contribute to improving community risk governance.

Finally, the results of the computational experiments point to promising avenues for improving the resilience-quantification framework. First, it can be enriched by modeling and monitoring a larger array of resilience metrics, some specific to community sectors and infrastructure systems components, others aggregated at the system level. Second, it can be generalized to model more complex repair planning and execution strategies (e.g. multiple repair crews), as well as to include the influence of other community infrastructure systems (e.g. transportation and communication) in the recovery modeling and simulation. Third, it can be expanded to include agent-based models of how disasters affect the population, such as casualties, evacuations and permanent or temporary relocations. Fourth, it can be made more sophisticated by refining the behavior of agents and modeling the interactions between them based on the instantaneous values of the monitored resilience metrics. Such a model could, with reasonable expectations, be calibrated against the damage absorption and recovery data collected after the past earthquake disasters.